\documentclass[preprint]{aastex}

\received{}
\accepted{}

\slugcomment{Submitted to \aj}

\author{
Gordon T. Richards\altaffilmark{2},
Michael A. Weinstein\altaffilmark{2},
Donald P. Schneider\altaffilmark{2},
Xiaohui Fan\altaffilmark{3},
Michael A. Strauss\altaffilmark{4},
Daniel E. Vanden Berk\altaffilmark{5},
James Annis\altaffilmark{5},
Scott Burles\altaffilmark{5,6},
Emily M. Laubacher\altaffilmark{6,7},
Donald G. York\altaffilmark{6,8},
Joshua A. Frieman\altaffilmark{5,6},
David Johnston\altaffilmark{6},
Ryan Scranton\altaffilmark{6},
James E. Gunn\altaffilmark{4},
\v{Z}eljko Ivezi\'{c}\altaffilmark{4},
R. C. Nichol\altaffilmark{9},
Tam\'as Budav\'ari\altaffilmark{10,11},
Istv\'an Csabai\altaffilmark{10,11},
Alexander S. Szalay\altaffilmark{11},
A. J. Connolly\altaffilmark{12},
Gyula P. Szokoly\altaffilmark{13},
Neta A. Bahcall\altaffilmark{4},
Narcisco Ben\'{\i}tez\altaffilmark{11},
J. Brinkmann\altaffilmark{14},
Robert Brunner\altaffilmark{15},
Masataka Fukugita\altaffilmark{16},
Patrick B. Hall\altaffilmark{4,17},
G. S. Hennessy\altaffilmark{18},
G. R. Knapp\altaffilmark{4},
Peter Z. Kunszt\altaffilmark{11},
D.Q. Lamb\altaffilmark{6},
Jeffrey A. Munn\altaffilmark{19},
Heidi Jo Newberg\altaffilmark{20},
Chris Stoughton\altaffilmark{5}
}

\altaffiltext{2}{Department of Astronomy and Astrophysics, The Pennsylvania State University, University Park, PA 16802}
\altaffiltext{3}{Institute for Advanced Study, Olden Lane, Princeton, NJ 08540}
\altaffiltext{4}{Princeton University Observatory, Princeton, NJ 08544}
\altaffiltext{5}{Fermi National Accelerator Laboratory, P.O. Box 500, Batavia, IL 60510}
\altaffiltext{6}{The University of Chicago, Department of Astronomy and Astrophysics, 5640 S. Ellis Ave., Chicago, IL 60637}
\altaffiltext{7}{The University of Dayton, Department of Physics, Dayton, OH 45469-2314}
\altaffiltext{8}{The University of Chicago, Enrico Fermi Institute, 5640 S. Ellis Ave., Chicago, IL 60637}
\altaffiltext{9}{Dept. of Physics, Carnegie Mellon University, 5000 Forbes Ave., Pittsburgh, PA-15232}
\altaffiltext{10}{Department of Physics, E\"otv\"os University, Budapest, Pf.\ 32, Hungary, H-1518}
\altaffiltext{11}{Department of Physics and Astronomy, The Johns Hopkins University, 3701 San Martin Drive, Baltimore, MD 21218}
\altaffiltext{12}{Department of Physics and Astronomy, University of Pittsburgh, Pittsburgh, PA 15260}
\altaffiltext{13}{Astrophysikalisches Institut Potsdam, An der Sternwarte 16, D-14482 Potsdam, Germany}
\altaffiltext{14}{Apache Point Observatory, P.O. Box 59, Sunspot, NM 88349-0059}
\altaffiltext{15}{Department of Astronomy, California Institute of Technology, Pasadena, CA 91125}
\altaffiltext{16}{University of Tokyo, Institute for Cosmic Ray Reserach, Kashiwa, 2778582, Japan}
\altaffiltext{17}{Pontificia Universidad Cat\'{o}lica de Chile, Departamento de Astronom\'{\i}a y Astrof\'{\i}sica, Facultad de F\'{\i}sica, Casilla 306, Santiago 22, Chile}
\altaffiltext{18}{U.S. Naval Observatory, 3450 Massachusetts Ave., NW, Washington, DC  20392-5420}
\altaffiltext{19}{U.S. Naval Observatory, Flagstaff Station, P.O. Box 1149, Flagstaff, AZ  86002-1149}
\altaffiltext{20}{Physics Department, Rensselaer Polytechnic Institute, SC1C25, Troy, NY 12180}

\begin{document}

\title{Photometric Redshifts of Quasars\footnote{Based on observations
obtained with the Sloan Digital Sky Survey.}}

\begin{abstract}

We demonstrate that the design of the Sloan Digital Sky Survey (SDSS)
filter system and the quality of the SDSS imaging data are sufficient
for determining accurate and precise photometric redshifts
(``photo-$z$''s) of quasars.  Using a sample of 2625 quasars, we show
that photo-$z$ determination is even possible for $z\le2.2$ despite
the lack of a strong continuum break that robust photo-$z$ techniques
normally require.  We find that, using our empirical method on our
sample of objects known to be quasars, approximately 70\% of the
photometric redshifts are correct to within $\Delta z = 0.2$; the
fraction of correct photometric redshifts is even better for $z>3$.
The accuracy of quasar photometric redshifts does not appear to be
dependent upon magnitude to nearly $21^{\rm st}$ magnitude in $i'$.
Careful calibration of the color-redshift relation to $21^{\rm st}$
magnitude may allow for the discovery of on the order of $10^6$
quasars candidates in addition to the $10^5$ quasars that the SDSS
will confirm spectroscopically.  We discuss the efficient selection of
quasar candidates from imaging data for use with the photometric
redshift technique and the potential scientific uses of a large sample
of quasar candidates with photometric redshifts.

\end{abstract}


\section{Introduction}

``Astronomical photometry is best considered as low-resolution
spectroscopy,'' said \citet{bes90}.  With carefully calibrated CCD
imaging data in five broad passbands ($u'$,$g'$,$r'$,$i'$,$z'$;
\citealt{fig+96}) that have sharp edges and little overlap, the
imaging survey of the Sloan Digital Sky Survey (SDSS;
\citealt{yor+00}) is designed to put this philosophy into practice.
For many applications the imaging part of the Survey can be treated as
an $R\sim4$ objective prism survey with wavelength coverage from
$\sim3000\,{\rm \AA}$ to $\sim10,500\,{\rm \AA}$.

In recent years, the practice of measuring redshifts of galaxies using
multiple band photometry has become both popular and powerful
\citep[e.g.,][]{bcs+97,cbs+99}, although the concept has been around
for quite some time \citep{bau62}.  The popularity of photometric
redshifts is not surprising given the impressive successes of the
method and the differences in exposure times between spectroscopy and
broad-band imaging.  The effectiveness of the method is primarily the
result of discontinuities in the spectral energy distribution (SED) of
galaxies, such as the $4000\,{\rm \AA}$ break.  A similar feature,
namely the discontinuity caused by the onset of the Lyman-$\alpha$
forest, allows one to accurately estimate redshifts for $z>3$ quasars
and galaxies (the Lyman-$\alpha$ forest is observable in the optical
by $z=2.2$, but it is not until $z=3$ that it makes the $u'-g'$ colors
red enough to distinguish them from lower redshift quasars).

Although the Lyman-$\alpha$ break allows for accurate redshift
determinations from photometry of $z>3$ quasars, the overall spectrum
of quasars is well-described by a power-law continuum, which is
invariant under redshift.  This means that, for quasars at redshifts
too small for the Lyman-$\alpha$ forest to be observable in the
optical from the ground, quasar photo-$z$ would be impossible if
quasar spectra were purely power-laws.  Despite the fact that quasars
also have emission line features that cause redshift-dependent color
changes that might be expected to help with low-$z$ ($z\le2.2$) quasar
photo-$z$ determinations, the photometric redshift technique has
hitherto not been effective for low-$z$ quasars because of three
effects: 1) this technique requires precision photometry --- errors in
the colors must be smaller than the structure in the color-redshift
relation, which is typically not possible (at these redshifts) with
photographic material; 2) filters that overlap significantly tend to
smooth out the features in the color-redshift relation; and 3) more
than two colors are needed to break the degeneracy (i.e., to
distinguish between two or more redshifts that have similar colors).

Of course, any color-selection of quasars is essentially a crude
attempt at determining photometric redshifts.  For example, the
selection of quasars as ultra-violet excess (UVX) point sources is
equivalent to predicting that the objects will turn out to be quasars
with $z<2.2$.  Similarly, very red outliers from the stellar locus
that are point sources are quite likely to be $z>3$ quasars.  However,
in this work, we go one step further and predict specific redshift
values for objects as opposed to predicting broad redshift ranges.

In \citet[hereafter R01]{ric01}, we showed that the colors of quasars
in the SDSS photometric system are a strong function of redshift for
$0 < z < 5$.  Furthermore, the colors of quasars at similar redshifts
have a relatively small dispersion.  We find that the existence of
significant structure in the color-redshift relation and the
reasonably small scatter in the SDSS colors at a given redshift allows
for the determination of photometric redshifts for quasars from the
SDSS photometry.

In this paper and in a companion paper \citep{bud01}, we present the
first large-scale determination of photometric redshifts for quasars,
at $0 < z < 5$, using only four colors derived from the five SDSS
broad-band magnitudes.  This result signals the beginning of practical
applications of quasar photometric redshifts that can be used for
scientific programs.  In this paper, we discuss an empirical quasar
photo-$z$ method that relies upon the structure that is present in the
empirical quasar color-redshift relation.  Photometric redshifts are
determined by minimizing the $\chi^2$ between the colors of a new
object and the colors of known quasars as a function of redshift.  In
\citet{bud01} we discuss the application of proven galaxy photo-$z$
techniques to quasars.  These techniques include other empirical
methods, such as nearest neighbor searches and polynomial fitting.  In
addition, \citet{bud01} discuss both a standard spectral template
fitting procedure and a hybrid approach that allows for the
construction of multiple templates.  It is our hope that further
analysis using the method presented herein and the methods presented
by \citet{bud01}, or a combination thereof, will result in a method
for determining photometric redshifts for quasars that rivals the
usefulness of the galaxy photometric redshifts.

This paper is structured as follows.  Section~2 gives a brief
description of the data used herein.  More details can be found in
R01, where this data set was defined.  In \S~3, we describe a simple
technique for determining quasar photometric redshifts, and
investigate how we might improve our photometric redshift predictions.
In \S~4 we discuss some science that can be done with a sample of
``quasars'' with photometric redshifts.  Finally, \S~5 presents our
conclusions.

\section{The Data}

The sample studied herein is the same sample analyzed in R01 and
includes 2625 quasars.  Of these 2625 quasars, 801 were previously
known quasars from the NASA Extragalactic Database (NED) that were
matched to objects in the SDSS imaging catalog.  A total of 1983 of
the 2625 quasars were independently discovered or recovered during
SDSS spectroscopic commissioning.  The photometry for these 1983
quasars was taken from the photometric catalogs of four scans of the
Celestial Equator that were taken between 1998 September 19 and 1999
March 22, which include data ``runs'' 94, 125, 752 and 756.  These are
2.5 degree wide equatorial scans that were observed and processed as
part of the commissioning phase of the SDSS imaging survey.  There is
considerable (but not complete) overlap between these 1983 SDSS
commissioning quasars and those used by \citet{van+01} in the
construction of a composite quasar spectrum using over 2200 SDSS
quasar spectra.

The photometry for all previously known quasars in the sample was
given in R01.  The photometry for the new SDSS quasars is available as
part of the SDSS Early Data Release; details regarding the new SDSS
quasars are given by \citet{sto+01}, including a discussion of the
selection algorithm used to target quasars during SDSS commissioning.
Quasars were selected primarily as outliers from the stellar locus;
the magnitudes have been corrected for Galactic reddening according to
\citet{sfd98}.  In addition to ``normal'' quasars, the sample includes
such objects as Broad Absorption Line (BAL) quasars, low-redshift
low-luminosity Seyfert galaxies, etc.  We note that the inclusion of
such objects is likely to produce results slightly different from
those results that we would obtain with a sample of purely ``normal''
quasars.

We emphasize that the quasar target selection algorithm underwent many
changes during the commissioning phase and that the SDSS quasars that
make up the majority of the sample studied herein in no way constitute
a homogeneous sample.  The final SDSS Quasar Target Selection
Algorithm, which differs somewhat from the preliminary algorithm
described by \citet{sto+01}, will be discussed by \citet{ric+01} and
\citet{new+01}.  For additional references and more details regarding
the entire data set (not just the SDSS commissioning objects), see
R01.

\section{Photometric Redshifts for Quasars}

The SDSS CCD mosaic imaging camera \citep{gcr+98} produces accurate
($\le 0.03$ mag) photometry that makes it possible to discern
structure in the quasar color-redshift relation.  As can be seen from
Figure~\ref{fig:fig1} (adapted from R01), where we plot the measured
colors of 2625 quasars (grey points) along with the median color as a
function of redshift (solid black curve), the color-redshift relation
for quasars in the SDSS filter set shows a relatively tight
distribution with a considerable amount of structure.\footnote{When
referring to actual measured values (such as in the Figures), the
filter names are designated as, for example, $r^*$ instead of $r'$, in
order to indicate the preliminary nature of the SDSS photometry
presented herein; however, we will typically use the $r'$ convention
whenever we are not discussing specific measurements.  The magnitudes
are given as asinh magnitudes \citep{lgs99}.}  Red outliers are
primarily anomalously red quasars (hereafter, referred to as
``reddened'' quasars, although is it not clear whether they are
actually reddened or simply redder than average), which will be
discussed in Section~\ref{sec:red}.  Also note the sharp changes in
color due to absorption by the Lyman-$\alpha$ forest as the quasars
move to higher redshifts.  With the four colors derived from the
five-band SDSS photometry, it is possible to break the degeneracy of
the color-redshift relation with sufficient accuracy and efficiency
that quasars can be assigned redshifts purely from their colors.

\subsection{The $\chi^2$ Minimization Technique}

\begin{figure}[p]
\epsscale{1.0}
\plotone{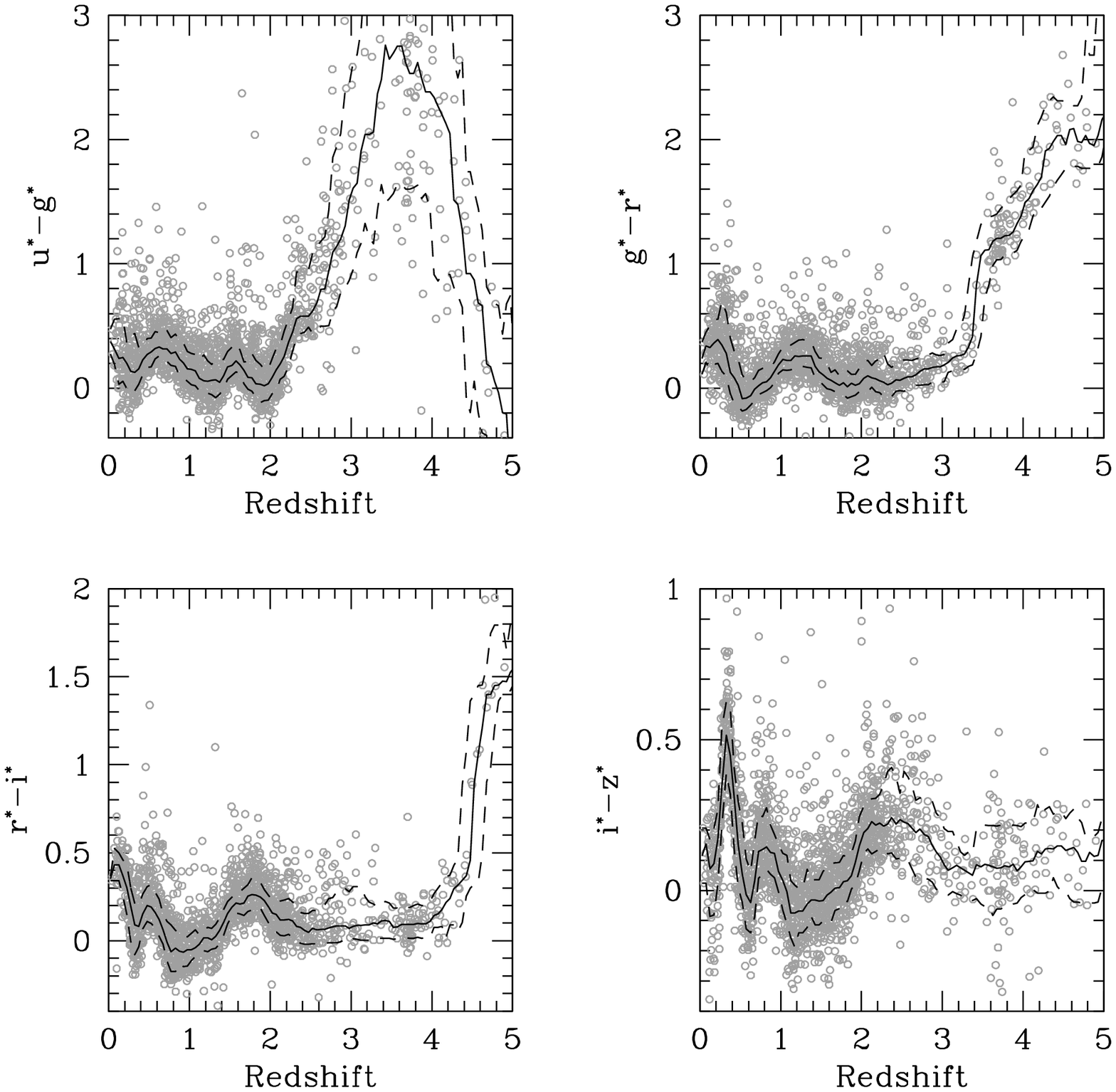}
\caption{SDSS colors versus redshift for 2625 quasars between $z=0$
and $z=5$ (grey points), the median color as a function of redshift
(solid black curve) and the 1-$\sigma$ error width in the
color-redshift relation (dashed black curve).  [Adapted from R01.]
\label{fig:fig1}}
\end{figure}

As a starting point for our photometric redshift investigation, we
begin with one of the simplest methods for determining photometric
redshifts given the structure that is seen in Figure~\ref{fig:fig1}.
First, we construct an empirical color-redshift relation based on the
median colors of quasars as a function of redshift.  The
color-redshift relation was formed by determining the median quasar
colors in bins with centers at 0.05 intervals in redshift, and widths
of 0.075 for $z\le2.15$, 0.5 for $z>2.5$, and 0.2 for $2.15 \le z \le
2.5$.  This relation is based on the assumption that, in our sample,
quasars at a given redshift have similar colors; this assumption does
not necessarily mean that the restframe color of all quasars should be
the same (i.e., there is not necessarily a redshift independent
spectral energy distribution).  Our empirical color-redshift relation
is shown by the solid black line in Figure~\ref{fig:fig1}; the colors
that encompass 68\% of the data in each redshift bin, thus defining an
effective $1\sigma$ error width in the color-redshift relation, are
given by the dashed black line.  The median and error vectors for this
data set are also computed in R01 and are given in Table~3 in that
paper.  Figure~\ref{fig:fig1} contains quasars that were identified by
R01 as a population of reddened quasars, which includes objects like
Broad Absorption Line (BAL) quasars; these anomalously red objects are
removed before computing the median colors as a function of redshift.

Photometric redshifts are then determined by minimizing the $\chi^2$
between all four observed colors and the median colors as a function
of redshift.  All colors are corrected for Galactic reddening
according to \citet{sfd98}.  The $\chi^2$ (for each redshift ---
indicated by the index, ``$z$'') is computed as
\begin{equation}
\chi^2_z = \frac{[(u'-g')-(u'-g')_z]^2}{\sigma_{(u'-g')}^2+\sigma_{(u'-g')_z}^2} + C_{gr} + C_{ri} + C_{iz},
\end{equation}
where $(u'-g')$ is the measured $u'-g'$ color of the object,
$(u'-g')_z$ is the color from the median color-redshift relation at a
given redshift, $\sigma_{(u'-g')}$ is the photometric error in the
$u'-g'$ color, which is given by
$(\sigma_{u'}^{2}+\sigma_{g'}^{2})^{1/2}$, and $\sigma_{(u'-g')_z}$ is
the $1\sigma$ error width of the median color-redshift relation as a
function of redshift.  Since the latter term is relatively independent
of redshift (see Figure~\ref{fig:fig1}), we have chosen to use a fixed
value for all redshifts.  The terms $C_{gr}$, $C_{ri}$, and $C_{iz}$
are analogous to the term for the $u'-g'$ colors, but for $g'-r'$,
$r'-i'$, and $i'-z'$, respectively.  The most probable redshift is
given by the redshift that produces the smallest value of $\chi^2$.
Slightly different formulations of $\chi^2$ may be more appropriate
depending on certain assumptions, such as whether the width of the
color distribution is independent of redshift or if magnitude errors
are correlated.

\begin{figure}[p]
\epsscale{1.0}
\plotone{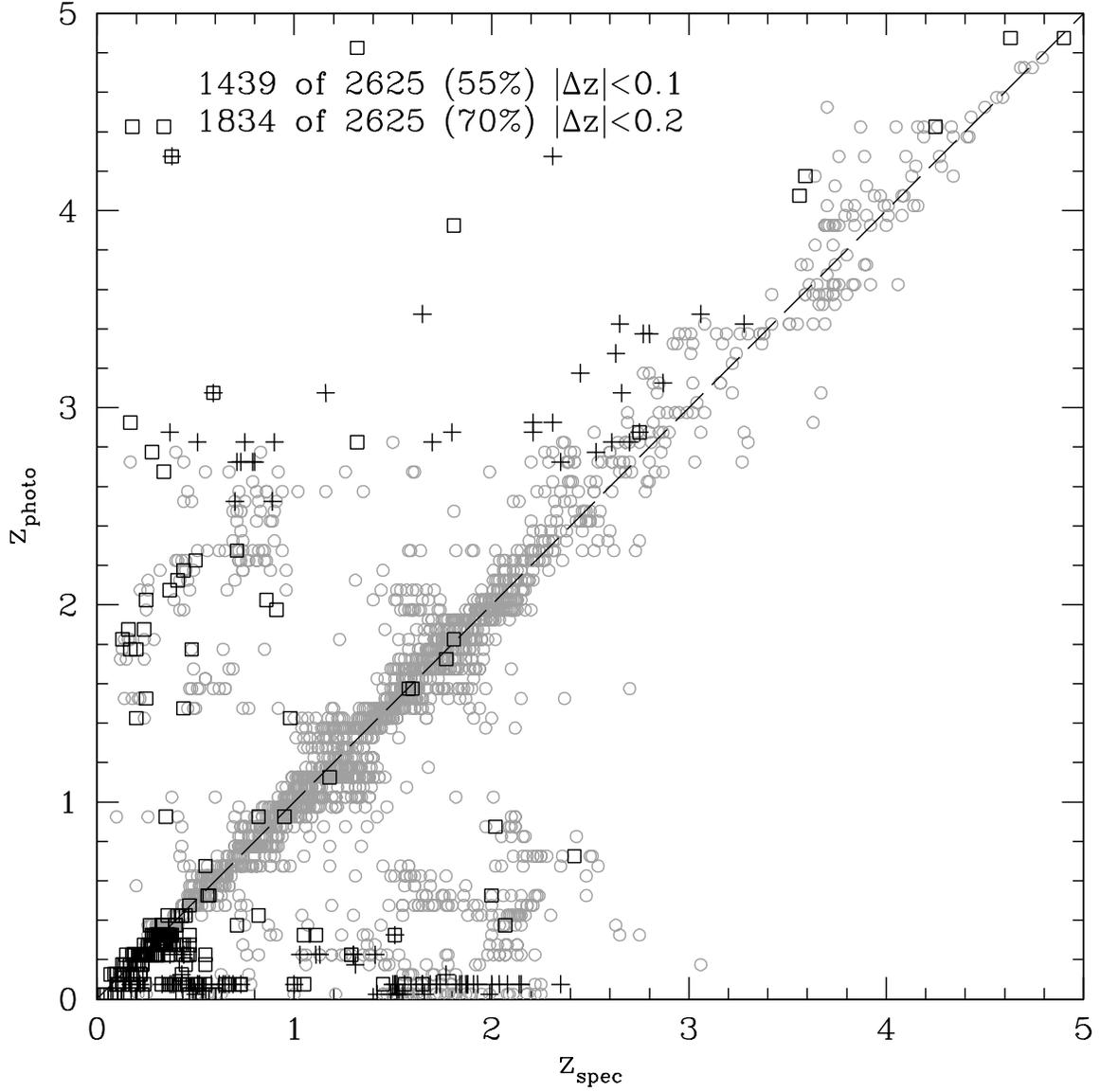}
\caption{Photometric redshift versus spectroscopic redshift for all
quasars in our sample.  Points marked with a black cross are anomalously
red quasars (see Section~\ref{sec:red}) and boxes indicate extended
sources (see Section~\ref{sec:extended}). \label{fig:fig2}}
\end{figure}

\begin{figure}[p]
\epsscale{1.0}
\plotone{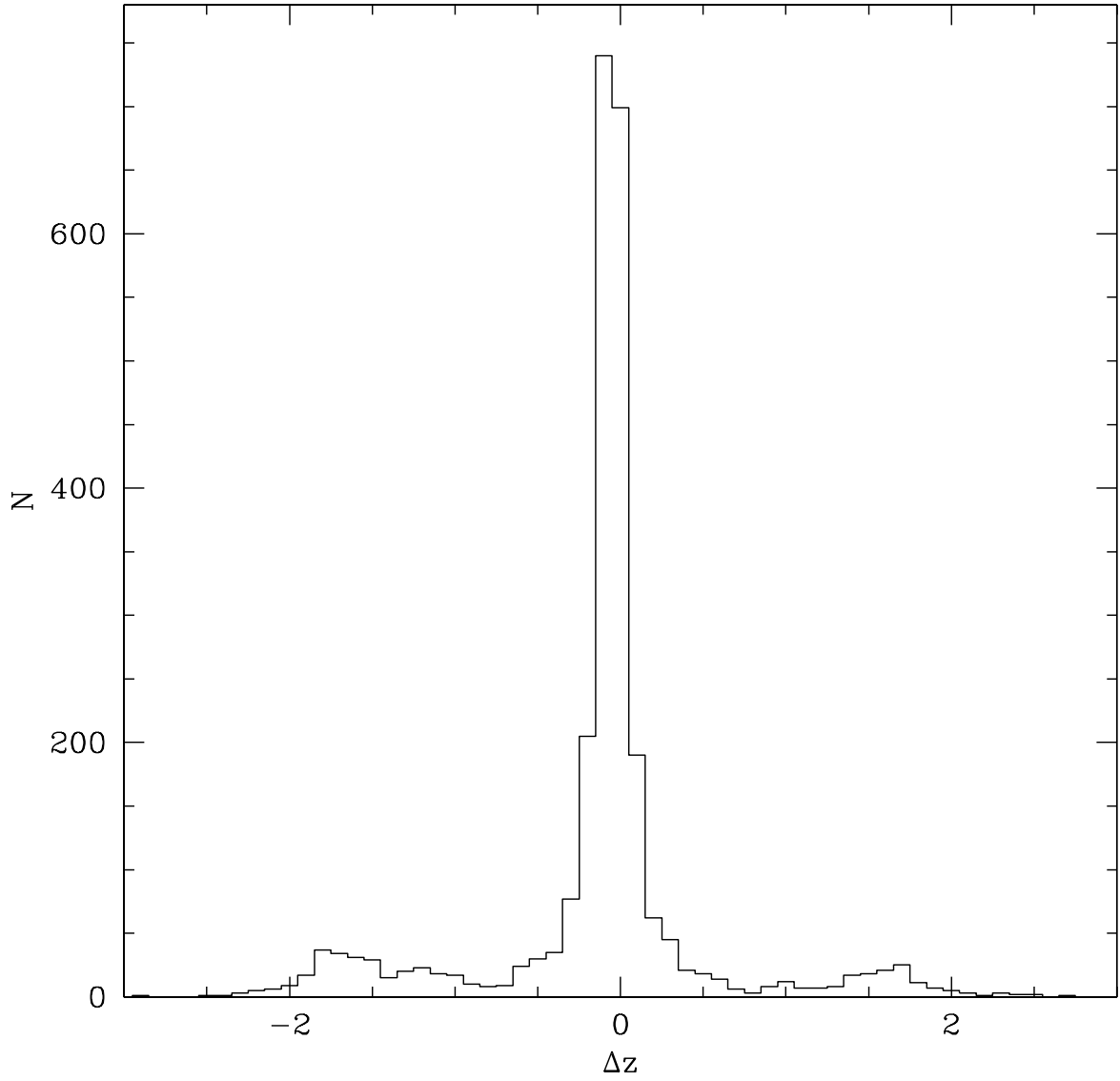}
\caption{Histogram of photometric redshift errors ($\Delta z =
z_{photo} - z_{spec}$) for the data from
Figure~\ref{fig:fig2}.\label{fig:fig3}}
\end{figure}

\begin{figure}[p]
\epsscale{1.0}
\plotone{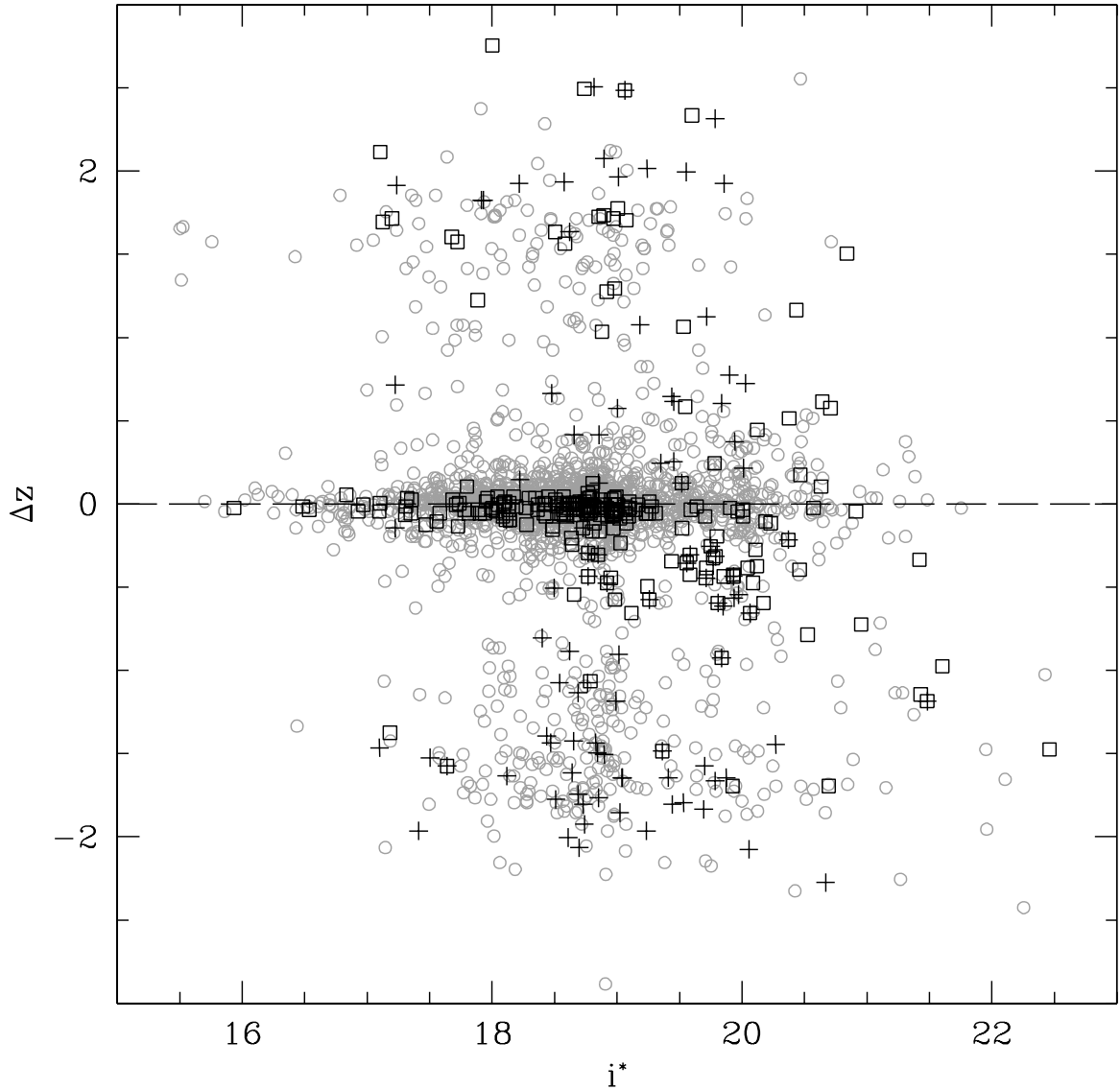}
\caption{Photometric redshift error ($\Delta z = z_{photo} -
z_{spec}$) as a function of SDSS $i^*$ magnitude.  There is no
significant deviation from an average redshift error of zero as a
function of magnitude.  Plot symbols are the same as for
Figure~\ref{fig:fig2}.\label{fig:fig4}}
\end{figure}

\begin{figure}[p]
\epsscale{1.0}
\plotone{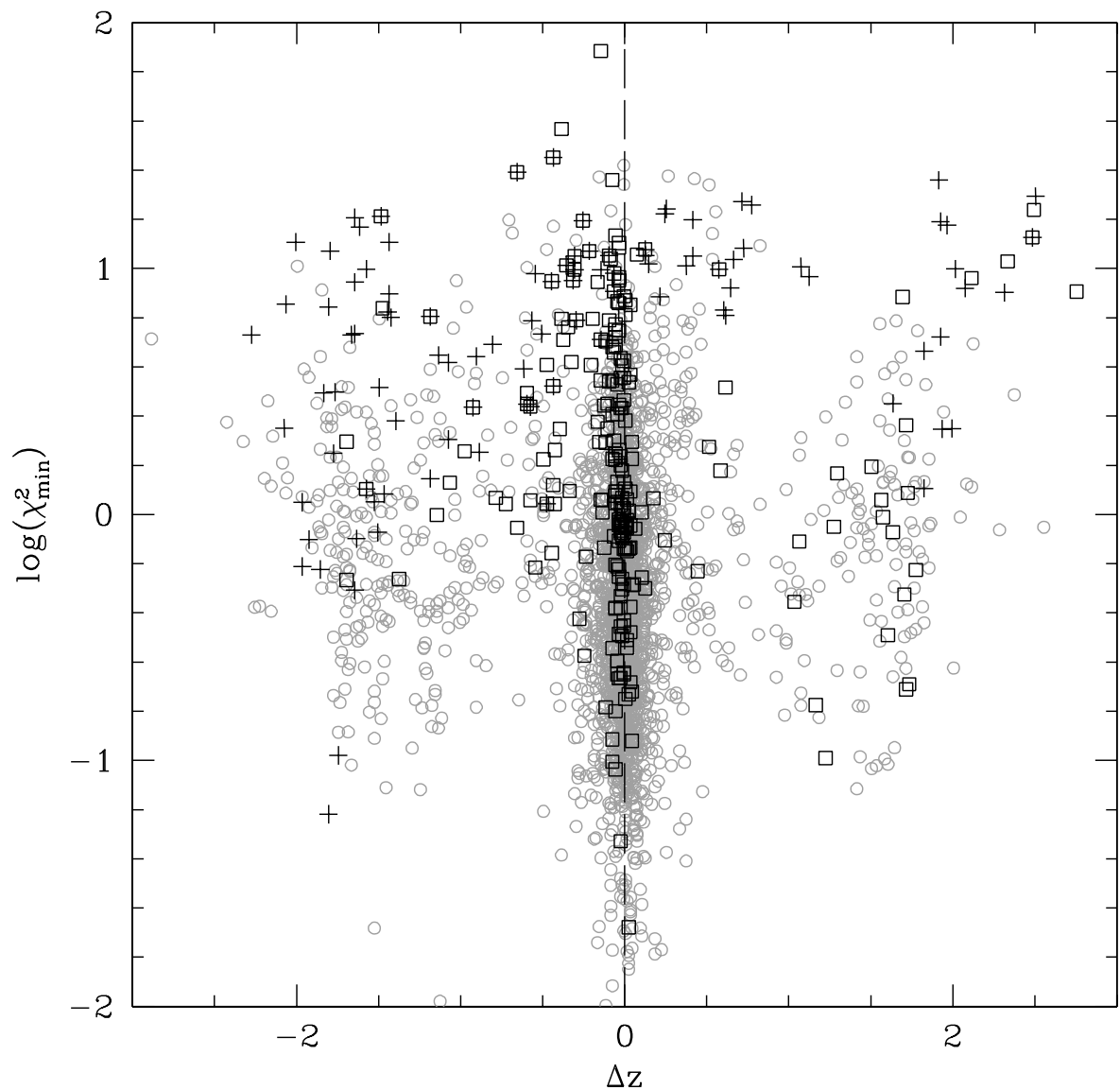}
\caption{Photometric redshift error ($\Delta z = z_{photo} -
z_{spec}$) as a function of minimum $\chi^2$.  Unfortunately, there is
no overall trend toward smaller minimum $\chi^2$ with decreasing
photometric redshift error.  Plot symbols are the same as for
Figure~\ref{fig:fig2}.\label{fig:fig5}}
\end{figure}

The first test of this photometric redshift technique is to determine
if it can recover the redshifts for the objects that comprise the
input data set.  The results of such an exercise are shown in
Figures~\ref{fig:fig2}~to~\ref{fig:fig5}, which display the results
for the 2625 quasars from R01.  In Figure~\ref{fig:fig2}, we plot the
predicted photometric redshift versus the spectroscopic redshift for
all of the quasars in our sample: the same quasars that went into
making the empirical color-redshift relation.  We find that 1834 of
2625 quasars (or 70\%) have photometric redshifts that are correct to
within $|\Delta z| = 0.2$ or better.  The RMS deviations from the
correct values are $\Delta_{{\rm all}} = 0.676$ and $\Delta_{0.3} =
0.099$, where $\Delta_{{\rm all}}$ is the RMS for the whole sample and
$\Delta_{0.3}$ is the RMS for those objects whose photometric
redshifts are within $|\Delta z| = 0.3$ of the spectroscopic redshift.
Considering $z>3$ alone, we see that our results are even better than
for all redshifts combined.  Points marked by crosses are those
quasars which are deemed to be anomalously red; they are the points in
Figure~\ref{fig:fig1} that lie well above the median relation (see R01
for discussion).  Points marked by squares are extended sources; the
colors of these objects may be affected by the light from the host
galaxy.

Further improvements to our method will come from understanding the
distribution of both extended and reddened quasars in addition to
investigating inherent redshift degeneracies, all of which are obvious
in Figure~\ref{fig:fig2}.  Sections~\ref{sec:red},~\ref{sec:extended},
and ~\ref{sec:degen} are devoted to discussions of how we might reduce
the fraction of objects with incorrect photometric redshift estimates.
It is also important to realize that all of the objects in our sample
are known to be quasars.  For actual scientific applications where we
know only the colors, obviously not all of the objects will be
quasars; this fact must be taken into account in the determination of
our efficiency, as we discuss in Section~\ref{sec:million}.

That a solid majority of the quasars have accurate photometric
redshifts can be seen more clearly in Figure~\ref{fig:fig3}, where we
plot a histogram of the differences between the photometric redshifts
and the spectroscopic redshifts.  This histogram reveals that the
photometric redshift error has a relatively Gaussian core with
$\sigma\sim0.1$ that contains $\sim70$\% of the quasars, and ``side
lobes'' at $\Delta z = \pm1.5$ that each contain about 15\% of the
objects.

Figure~\ref{fig:fig4} shows $\Delta z$ as a function of the quasar
$i^*$ magnitude.  That there is no significant deviation from zero as
a function of magnitude bodes well for using the photometric redshift
technique to estimate redshifts for quasars fainter than the limit of
our sample (until the magnitude errors become too large).  The
potential value of this result for creating a large sample of faint
quasars is discussed in Section~\ref{sec:million}.

Finally, in Figure~\ref{fig:fig5}, we plot $\Delta z$ versus the
minimum value of $\chi^2$ for each quasar.  As long as our photometric
errors are relatively small, one would expect the minimum $\chi^2$ to
decrease with increasing photometric redshift accuracy; if this were
true, it would provide an estimate of the accuracy of the photometric
redshift.  Unfortunately, we see no such trend (except for the very
smallest values of $\chi^2$).  As a result of color degeneracies as a
function of redshift, an erroneously predicted redshift can have a
small value of $\chi^2$.  The value of $\chi^2$ can be low for more
than one redshift, and the lowest $\chi^2$ value --- which may not be
significantly lower than the others --- can be associated with an
incorrect redshift.  See Section~\ref{sec:degen} for a more detailed
discussion of redshift degeneracies.
 
\begin{figure}[p]
\epsscale{1.0}
\plotone{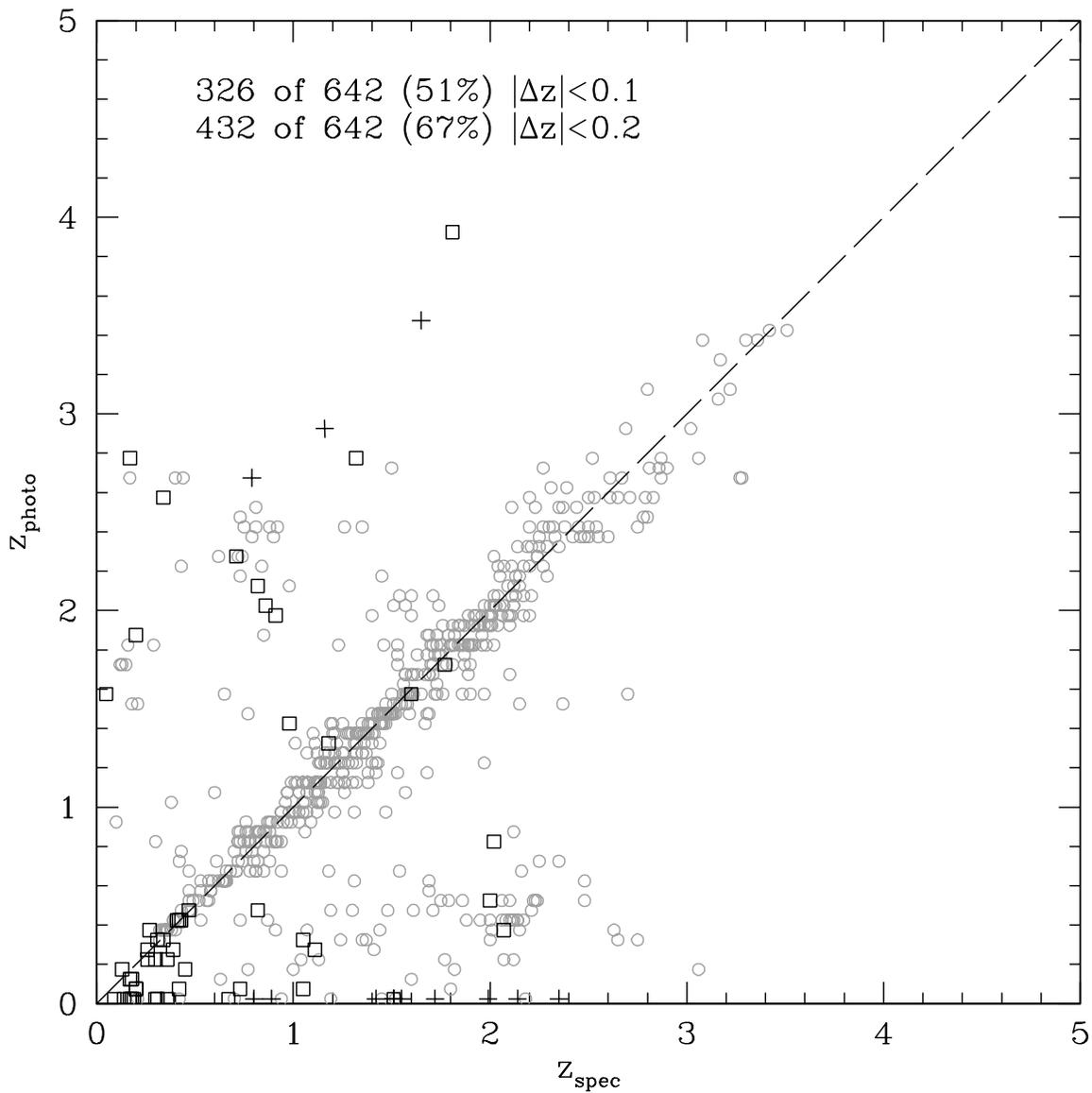}
\caption{Photometric redshifts versus spectroscopic redshift for NED
quasars using only SDSS spectroscopic commissiong quasars to compute
the median color-redshift relation.  Plot symbols are the same as for
Figure~\ref{fig:fig2}.\label{fig:fig6}}
\end{figure}

A more challenging test than the analysis presented in
Figures~\ref{fig:fig2}~to~\ref{fig:fig5} would be to apply this method
to a set of quasars that is distinct from the set used to determine
the median color-redshift values; Figure~\ref{fig:fig6} shows the
results of such an exercise.  Here we have calculated photometric
redshifts for the subsample of SDSS quasars that were previously known
from the NASA Extra-Galactic Database (NED) [see R01 for more
details], using a color-redshift relation generated solely from those
quasars in our sample that were not previously known to NED.  Once
again, the photometric redshifts of nearly 70\% of the input quasars
are correct to within $|\Delta z| = 0.2$.  The RMS values for this
sample are $\Delta_{{\rm all}} = 0.702$ and $\Delta_{0.3} = 0.105$,
essentially identical to the values above.  This result is extremely
encouraging; one would expect that the accuracy of our initial study
(e.g., Figure~\ref{fig:fig1}) would be an upper limit to the overall
accuracy of the technique and that using independent samples for the
median color-redshift relation and the test objects might produce
considerably worse results.  This is consistent with the statement
that the color-redshift relation of quasars is the same for both NED
and SDSS quasars.  As an additional test, we have also matched the 2dF
10k QSO catalog \citep{csb+01} to the SDSS catalog and have determined
photometric redshifts for a sample of 2dF quasars using SDSS
photometry.  We find that 1492 of our 2262 2dF/SDSS matches (all of
which have $z<3.2$) have photometric redshifts accurate to within
$|\Delta z| = 0.2$ --- yielding a similar accuracy (66\%) to our main
sample.

Our results are probably, to some extent, a reflection of how similar
the SED's of quasars are to one another.  That quasars are roughly
similar can be seen by comparing the quasar composites of
\citet{fhf+91}, \citet{bro00}, and \citet{van+01}, which were made
from quasars with different selection criteria, yet are still very
similar. Also, if there were a broader distribution of spectral
indices and/or emission line strengths, then the color-redshift
relation would show much less structure and it would not be possible
to determine photometric redshifts for $z<2.2$ quasars.
\citet{fhf+92} found that that 75\% of the variance in quasar spectra
can be accounted for by the first three principal components of their
composite spectrum.  The first component represents the quasar
emission line spectrum, whereas the second component includes the
power-law continuum slope and modifications to the emission lines.
These two components probably account for nearly all of the structure
and the width of the color-redshift relation as shown in
Figure~\ref{fig:fig1}.  As a result of these similarities, quasar
photometric redshifts are possible given sufficiently high-quality
photometry.

For the sake of comparison, \citet{hmp00}, using two samples of 51 and
46 quasars with photometry from 6 broad-band filters, were able to
determine photometric redshifts to within $|\Delta z| < 0.1$ for 23
(45\%) and 17 (37\%) of their quasars in each sample, respectively.
Similarly, \citet{wmr+01} report an accuracy of 50\% ($|\Delta z| <
0.1$) in their sample of 20 objects using 11 intermediate- and
broad-band filters.  Our accuracy (fraction of objects within $|\Delta
z| < 0.2$) is competitive with, if not considerably better than, these
previous results using a sample that is more than 50 times as large.
Furthermore, it is significant that our study uses only four
broad-band colors and that no effort beyond routine SDSS operations is
needed to obtain the required measurements.  

Thus far, we have applied our algorithm without any pre- or
post-processing of the data in the input sample.  However, it is clear
that an application of some constraints on the input sample will lead
to significant improvements in our results without introducing any
significant biases.  We now turn to a discussion of these techniques.

\subsection{Reddened Quasars\label{sec:red}}

Approximately 4\% of the quasars in our sample are substantially
redder than can be expected from Galactic extinction (R01).  Such
quasars present a problem for photometric redshift determination,
since their colors are significantly different from the intrinsic
colors of the average quasar at the same redshift.  For example, in
Figures~\ref{fig:fig2}, \ref{fig:fig4}, \ref{fig:fig5}, and
\ref{fig:fig6} we have marked quasars that are anomalously red for
their redshift with black crosses; many of these anomalously red
quasars are Broad Absorption Line (BAL) quasars.  As one expects, most
of the reddened quasars (92\%) have incorrect photometric redshifts;
moreover, an appreciable fraction (12\%) of the quasars with bad
photometric redshifts are reddened.

There are a number of ways to deal with reddened quasars to improve
the effectiveness of the photo-$z$ algorithm.  One could attempt to
recognize reddened quasars a priori and apply a correction to their
colors.  However, this process would only work if the red quasars were
clearly identifiable and could be corrected with some sort of global
correction factor; such a procedure is unlikely to be very effective
due to the broad range of reddening that is observed (the colors of
reddened quasars at a given redshift are consistent with the colors of
reddened quasars at nearly all other redshifts).  One could also
attempt to recognize reddened quasars {\em before} attempting to
calculate their photometric redshifts and then reject them from
consideration (as opposed to recognizing them and trying to make a
correction).

Identifying red quasars without knowing their redshifts may not be as
difficult as it might initially appear.  For example, the vast
majority of $z<2.2$ quasars have $u'-g' \le 0.6$, so a quasar
candidate with $u'-g' > 0.6$ is probably either a high-redshift
quasar, or a reddened low-redshift quasar (if the object is indeed a
quasar).  Similar approaches can be tried with outliers in the other
colors.  Of course, there can be redshift biases inherent in this
process: as can be seen in Figure~\ref{fig:fig1}, a redshift $1.6$
quasar that is reddened by 0.5 mag stands out more clearly as a red
outlier in $u'-g'$ than does a similarly reddened redshift 1.9 quasar,
as a result of the structure in the quasar color-redshift relation.

We have made a preliminary attempt to reject reddened quasars and to
study the effects of their rejection on our results.  Using only point
sources, we have found a series of color-cuts (based initially on the
$u'-z'$ colors) that allow for the identification of reddened quasars
that is roughly independent of redshift and unbiased against high-$z$
quasars.  These color-cuts are applied uniformly to the entire data
set --- without the knowledge of the actual redshifts of the quasars.
We find that the application of these color-cuts to reject reddened
quasars improves our (fractional) success rate to roughly 73\%.  In
order to apply this technique with confidence in the future, we need
more than the 40 quasars with $z>4$ that are in our current sample.
These additional $z>4$ quasars will determine the empirical scatter in
the colors for the highest redshift quasars, which, in turn, will
allow a better separation between real high-$z$ quasars and reddened
lower redshift quasars.

\subsection{Extended Objects\label{sec:extended}}

Extended sources, which are plotted as boxes in the figures, pose an
additional problem.  At high redshift ($z>1$), about half of the
quasars which are labeled as extended in the sample presented herein
are actually classified as point sources by the latest version of the
SDSS image processing software; such objects should not be a problem
in the future.  However, at low redshift there are low-luminosity
objects (Seyferts; see R01) whose colors are likely to be contaminated
by the light from the quasar host galaxy and will appear redder than
normal.  As with the reddened quasars discussed above, we must handle
such objects properly to get the best photometric redshifts.  At
first, it may be wise to simply exclude all extended quasars from our
median color-redshift relation and not attempt to determine
photometric redshifts for extended sources.  After all, it is clear
that these have relatively small redshifts (for quasars), which may be
sufficient constraint for most science applications.  However, in the
future, we would hope to be able to determine photometric redshifts
for these objects as well.

For our present analysis, we have included extended objects, which has
two effects.  First, it causes the median colors to be slightly redder
at low redshift than they would otherwise be.  Second, this means that
some extended objects will have quite good photometric redshifts, but
that very blue point sources at low redshift might have poorer
photometric redshifts than would otherwise be expected.  We expect
that those extended sources with poor photometric redshifts will tend
to have large photometric redshifts, since their redder colors will
push them to higher apparent redshifts.  A preliminary investigation
found that excluding extended objects from the median color-redshift
relation does not, in fact, improve our results for point sources;
however, more work is needed on the subject.

\subsection{Redshift Degeneracies\label{sec:degen}}

From Figure~\ref{fig:fig2} it is clear that there are some
degeneracies in the color-redshift relation.  Such an effect is
indicated by outliers that are symmetric around the
$z_{spec}=z_{photo}$ line.  For example, a number of $z_{spec}\sim2$
quasars are being assigned photometric redshifts near $0.4$; similarly
many $z_{spec}\sim0.4$ quasars are given photometric redshifts around
$2$.  To some extent, we may be unable to resolve these degeneracies;
however, we have reason to be encouraged that we may be able to
determine the correct redshifts for these objects.

\begin{figure}[p]
\epsscale{1.0}
\plotone{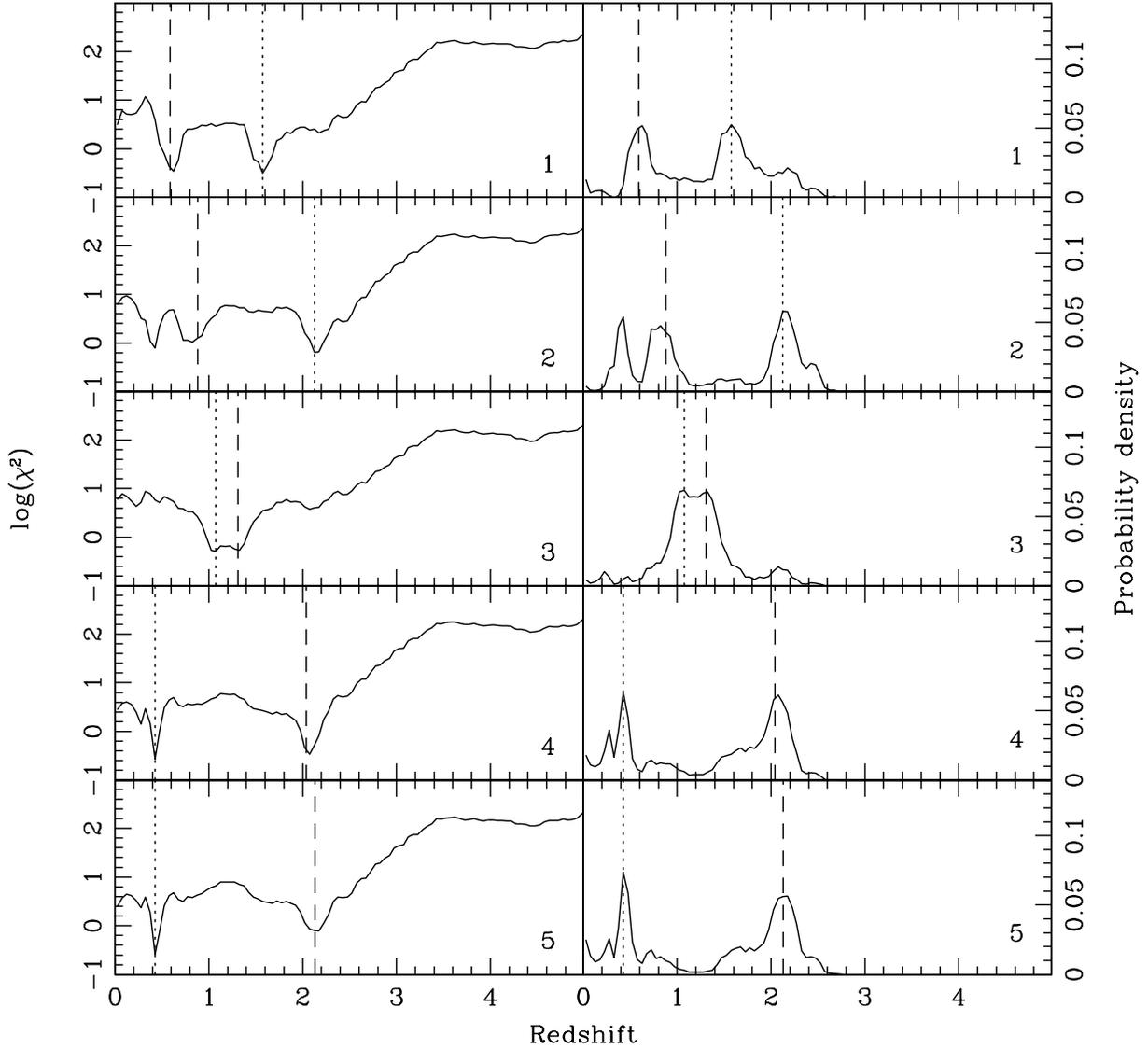}
\caption{Photometric redshift $\chi^2$ (left panels) and probability
density (right panels) values as a function of redshift for five
quasars with discrepant photometric redshifts.  Panels at the same $y$
location (which also have the same number in the lower right hand
corner) represent the same quasar.  The dashed line denotes the true
redshift, whereas the dotted line gives the most probable redshift
based on the $\chi^2$ minimization.\label{fig:fig7}}
\end{figure}

In Figure~\ref{fig:fig7}, we plot the $\chi^2$ (left panels) and the
probability density (right panels, see \S~\ref{sec:prob}) as a
function of redshift for five sample quasars whose spectroscopic and
photometric redshifts disagree by more than $0.2$.  Of particular
interest is that these objects (that were selected to illustrate the
following point) all have local minima at or near the true redshift.
As such, it may be possible to assign a second most probable redshift
to these objects that is much closer to the true redshift.  We find
that 331 of 791 objects (42\%) whose strongest minima produce
erroneous photometric redshifts have the correct redshift as a
secondary minimum.  If we can find a way to recognize such cases, then
our efficiency would improve to about 82\%.

We find that, for some objects, the minimum in $\chi^2$ for the {\em
wrong} redshift is narrower than the minimum for the correct redshift
(see quasars 4 and 5 in Figure~\ref{fig:fig7}).  This effect may be
due to objects with redshifts near the correct redshift having more
similar colors than objects with redshifts near the incorrect
redshift.  Allowing the algorithm to take the ``width'' of the minima
(i.e., the area under the curve) into account may improve our results.
In addition, we note that there are some cases where the minimum is
very broad or where there are two closely spaced minima and the
photometric redshift estimate is actually quite good (such as for
quasar 3 in Figure~\ref{fig:fig7}) despite the fact that it is off by
more than $0.2$ in redshift.

\begin{figure}[p]
\epsscale{1.0}
\plotone{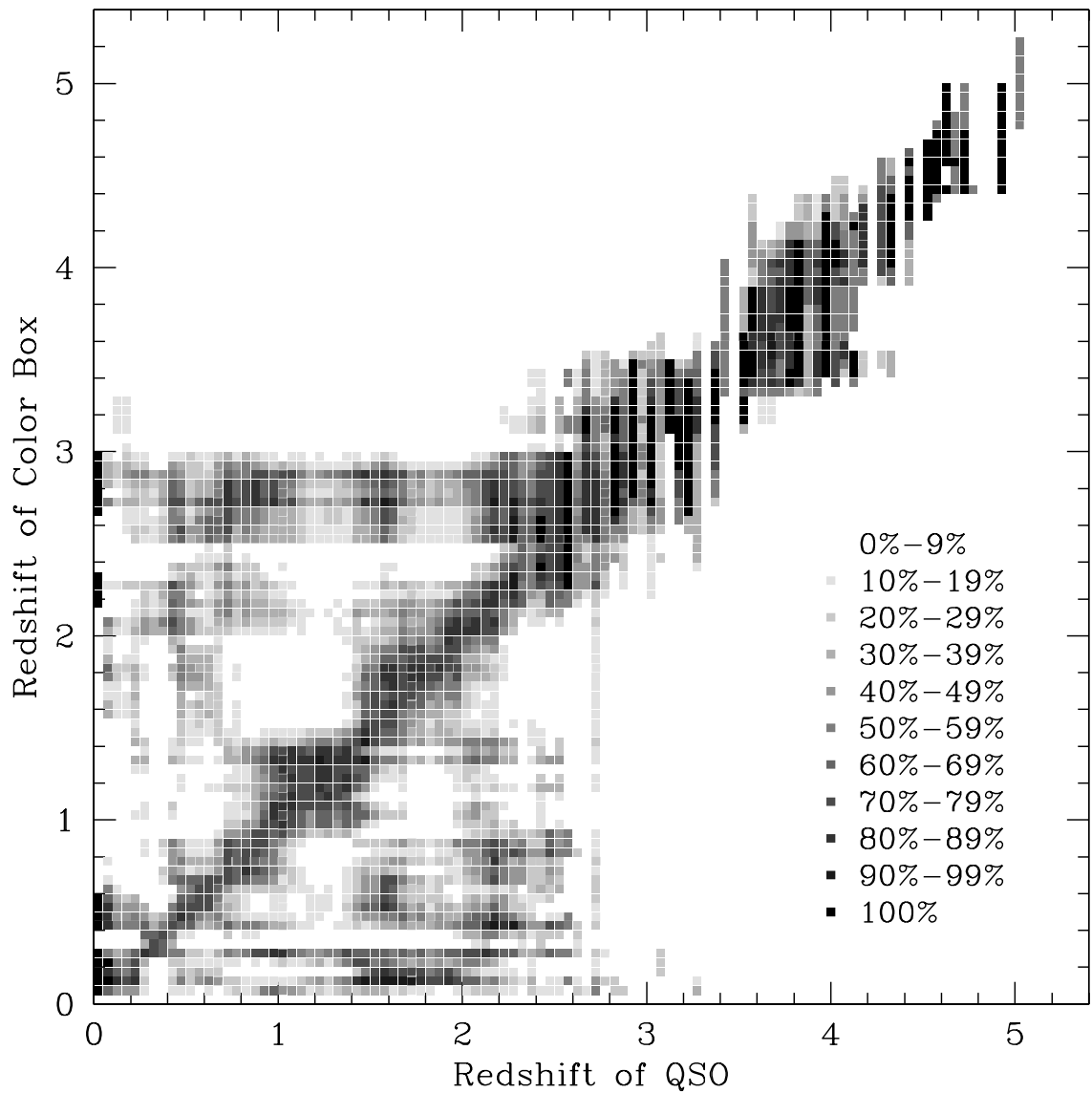}
\caption{Redshift degeneracy.  The grayscale of the pixel at
coordinates $(X,Y)$ indicates the fraction of quasars with redshift
$X$ which are within the 95\% color-box associated with redshift $Y$.
(See text for details.) \label{fig:fig8}}
\end{figure}

In order to give the reader an impression for which pairs of redshifts
are degenerate in our data, we have created Figure~\ref{fig:fig8}.  We
generate the figure in the following way.  First, we calculate the
95\% confidence interval about the median in each color, for redshift
bins of width 0.05.  For each redshift bin, we define a
four-dimensional box in color-space, whose lengths along each
dimension are the 95\% confidence intervals for $u'-g'$, $g'-r'$,
$r'-i'$, and $i'-z'$.  These boxes are called the ``95\%
color-boxes'', where ``95\%'' simply refers to the confidence
intervals that generated the box; in general, it will not be true that
95\% of the quasars in a redshift bin will lie within the color-box
associated with that redshift bin.

For the purposes of Figure~\ref{fig:fig8}, a quasar is considered to
be ``consistent'' with a certain redshift bin if its position in
color-space lies within the 95\% color-box for that redshift bin.  The
shade of gray of the point $(X,Y)$ represents the percentage of
quasars in redshift bin $X$ which lie within the 95\% color-box of
redshift bin $Y$.  Note that the fraction in each bin is independent
of the other bins; the sum along a column or row can be greater than
100\%.  One also must keep in mind that some of the redshift bins for
$z > 2.5$ contain very few ($<10$) quasars, so that the percentages
associated with these bins are not always terribly meaningful.

As one would expect, there are high percentages along the ``main
diagonal'' (the line $Y=X$).  This just means that most quasars with
redshift $X$ are ``consistent'' with having a redshift of $Y=X$.
However, high percentages can also be found at other regions of the
figure, for example: at approximately $(2.2,0.4)$, $(1.6,0.2)$,
$(1.6,0.7)$, $(2.2,0.7)$, and (to a lesser extent) their symmetric
counterparts across the main diagonal; and at a number of regions with
$Y \approx 2.8$ and $X < 2.8$.  These ``off-diagonal'' redshift pairs
are due to the degeneracies in the median color-redshift relation.  In
some cases these degeneracies may be caused by emission features
falling between two filters.  For example, $z=0.4$ and $z=2.2$ are
somewhat degenerate, perhaps in part because the bulk of the emission
from \ion{Mg}{2} and Lyman-$\alpha$, respectively, is emitted at
wavelengths that fall in the relatively unsensitive region between the
$u'$ and $g'$ filters.

One observes that Figure~\ref{fig:fig8} is not perfectly symmetric
about the main diagonal.  This is because the two axes are not on
equal footing: the $x$-axis deals with the actual distribution of
quasars in color-space, while the $y$-axis approximates those
distributions (and not very accurately) with the 95\% color-boxes.
Presumably, Figure~\ref{fig:fig8} would become more symmetric if the
color-box for each redshift bin was replaced by a region that more
closely resembled the actual distribution in color-space of the
quasars belonging to that redshift bin.

In some sense, Figure~\ref{fig:fig8} is a diagnostic for our
photometric redshifts.  For example, if the photometric redshift of a
quasar is around 2.2, we should be aware of the possibility that the
spectroscopic redshift is actually near 0.4, since quasars at both
redshifts are found in essentially the same region of color-space.
Note that, as a result of this diagnostic capability,
Figure~\ref{fig:fig8} resembles Figure~\ref{fig:fig2}.  Knowledge of
the degenerate regions of color-space, coupled with other information
(such as the expected redshift distribution of the sources) will help
to break the redshift degeneracies present in our data and will
improve our photometric redshift measurements.

\subsection{Future Work\label{sec:improve}}

The goal of this paper is not just to describe a photometric redshift
algorithm for quasars, but to demonstrate the ability to determine
photometric redshifts for quasars using SDSS imaging data --- a feat
which is significant in and of itself.  We are certainly encouraged
that our simple photo-$z$ method produces such accurate results;
however, the method produces an incorrect redshift about 30\% of the
time.  We have already discussed some specific changes that can be
applied to the basic algorithm to improve our efficiency and have
shown that it may be possible to increase our efficiency to nearly
85\% accuracy.  In the following sections, we introduce some less
specific and/or more time consuming possibilities for improving our
accuracy that are beyond the scope of this paper.

Most importantly, if we are to apply this technique effectively to do
real science, we must also be able to select quasars efficiently from
their colors alone.  Our true accuracy is the efficiency of the quasar
color-selection algorithm (roughly 70\%) multiplied by our photometric
redshift accuracy (roughly 70\%), yielding a true accuracy of more
like 50\%.  Thus it is clear that in addition to the need to improve
our photometric redshift accuracy, we must also find ways to improve
our quasar selection efficiency.

\subsubsection{More Data}

Clearly as more SDSS imaging data and spectroscopic redshifts are
acquired, we can refine our color-redshift relation, which should
allow for increased accuracy in determining photometric redshifts ---
particularly for $z>2.2$ where there are currently very few quasars in
each redshift bin.  For real $z>2.2$ quasars we already do quite well;
however, as we discussed above, additional high-$z$ quasars will help
us to discern between lower redshift quasars that are reddened and
normal high redshift quasars.  

Additional photometric data will allow us to reject more outliers when
making the color-redshift relation and will help us to understand how
to recognize these outliers --- without knowing their spectroscopic
redshifts.  We will further benefit by being able to better determine
the width of the color-redshift relation as a function of redshift and
whether that width is purely due to a range of optical spectral
indices, or other factors.  Better knowledge of the width of the
distribution will, in turn, allow for better understanding of what
redshifts have color degeneracies.  The combination of these
improvements in our knowledge may help to improve the accuracy of our
photometric redshifts.

\subsubsection{More Sophisticated Photo-$z$ Techniques}

Since the idea of photometric redshifts has been around for so long
\citep{bau62} and since the technique has been successful using the
Hubble Deep Fields \citep[e.g.,][]{sly97}, there is an extensive body
of literature that details a number of different techniques from which
we might be able to learn
\citep[e.g.,][]{bau62,koo85,ls86,ccs+95,lyf96,sly97,bcs+97,wbt98,cbs+99,bsccd00,ccs+00}.
As such, in addition to our simplistic $\chi^2$ minimization method,
we contemplate the use of other methods that have been tested on
galaxies.

In general, traditional galaxy techniques will be difficult to carry
over to quasars.  Unlike galaxies, the colors of quasars do not occupy
a relatively flat plane in color space and a given color does not
necessarily correspond to a unique redshift.  That is, $z$ is not
strictly a {\em function} of color.  As a result, PCA-type redshift
decompositions \citep{ccs+95} are probably not possible for quasars
(at least in terms of determining photometric redshifts).  For
high-redshift quasars where the color reddens rapidly as a result of
Lyman-$\alpha$ and Lyman-limit absorption, the color-redshift relation
is more similar to that of galaxies and traditional galaxy techniques
may be more applicable to quasars.

One might, a priori, think that SED-fitting techniques would also be
difficult for quasars, since the quasar spectral energy distribution
is very close to a power-law: unlike the SED for galaxies.  However,
in \citet{bud01} we show that quasar photo-$z$ is indeed possible
using SED-fitting methods with both a single SED template and a set of
four templates.  SED-fitting techniques can be helpful when attempting
to discriminate between different galaxy spectral types.  For quasars,
while there are clearly different types of objects (such as Broad
Absorption Line QSOs, BL Lacs, etc.), they are not typically divided
into spectral types in the same way that galaxies are divided into
spectral types.  This situation is partially due to a relative dearth
of uniform quasar spectra, thus we are hopeful that the situation will
change in the near future and that spectral typing of quasars will
become more common.

In addition, knowing the apparent magnitudes of the objects is not as
useful for quasars as it is for galaxies.  For galaxies, the redshift
is roughly correlated with apparent magnitude; more distant galaxies
tend to be fainter.  While it is also true that more distant quasars
tend to be fainter than closer quasars, the dynamic range of quasar
luminosities is so large that any potentially useful correlation is
strongly diluted.  However, this does bring up the fact that in using
only four colors, we are not using all of the information available,
and we are certainly free to add the observed brightness to our
equations.

There may be other approaches to be explored that may improve our
results.  One example would be to use a Bayesian-type analysis
\citep{ben00}, where we would weight our results based on the redshift
distribution and/or luminosity function of the quasars in our sample.
Such a weighting could help break the degeneracy in cases where two or
more photometric redshifts are equally likely.  Our initial attempts
at using the magnitude distribution as a Bayesian prior suggest that
an improvement of 3.5\% might be realized by applying this technique.

\subsubsection{Infra-red Data}

Additional improvements may come from adding infra-red (IR) colors to
our analysis; additional colors will contribute to the accuracy of the
method.  2MASS \citep{sss+97} provides IR imaging data for the entire
sky and makes an excellent sample to use for this application.  The
addition of 2MASS quasars \citep{bh01} to our analysis would be very
interesting; however, the overlap between the SDSS and 2MASS in
magnitude space is limited to the very brightest SDSS quasars
(typically, $r' \le 17.5$).  Since the SDSS will obtain spectra for
all of these objects, this exercise is not as useful as it might
otherwise be.  Nevertheless, there are applications for such an
analysis, but they are beyond the scope of this paper.

\subsubsection{Photometric Redshift Probabilities \label{sec:prob}}

In addition to finding more than one probable redshift for each
object, it would be desirable to give some measure of how confident
one can be that the redshift is correct.  As a preliminary attempt to
assign probabilities to our photometric redshifts, we define the
quantity $N e^{-\chi_{z}^2/2}$ to be an estimate of the probability
density as a function of redshift for each object, where the value of
$N$ normalizes the probability density to unity over the entire
redshift range.  Examples are given in the right hand panels of
Figure~\ref{fig:fig7} where we show the probability density functions
for the same quasars as in the left hand panels.  This presentation
emphasizes the most probable regions of redshift space.

We should also stress that, for some science applications, one will
not simply want the most probable redshift.  Rather, one might want to
know things like ``what is the probability that this object really has
a redshift larger than X'', or ``how likely is it that the correct
redshift is other than the most probable redshift''.  Clearly, the
technique presented herein allows for such applications; work on this
subject is in progress.

\subsection{Towards One Million Quasars\label{sec:million}} 

More than one hundred thousand SDSS quasar candidates brighter than
$i'\sim19$ will eventually have SDSS spectra.  Clearly, we can also
determine photometric redshifts for these quasars; however doing so
has limited value since these objects will already have spectroscopic
redshifts.  The ability to determine photometric redshifts from
$u',g',r',i',z'$ outside of the SDSS survey area will certainly be of
interest, but the true value in this method will come from photometric
redshift determination of faint objects for which spectroscopic
redshifts would require large amounts of telescope time.

In theory, we could determine photometric redshifts for as many as one
million quasars in the SDSS area {\em if} we can determine photometric
redshifts for objects as faint as $g'=21$ {\em and} if we can select
quasar candidates with very high efficiency.  We now turn to a
discussion of these issues.

\subsubsection{Photo-$z$ for Faint Quasars}

In principle, we could already determine photometric redshifts for a
large sample of faint quasars by using our existing data; however, in
practice, it would be better to get more spectra of fainter quasars in
order to calibrate the photometric redshifts at the faint end (i.e.,
to determine whether or not faint quasars have the same intrinsic
colors as bright quasars at the same redshift).  At least three
sources of faint quasar data are possible: 1) the SDSS Southern
Survey, where the SDSS will probe to fainter limits that in the
Northern Survey, 2) matching SDSS photometry to spectroscopically
confirmed quasars from the 2dF QSO Redshift Survey \citep{csb+01}, and
3) a separate faint quasar follow-up project.  To some extent, the
need for a test sample of fainter quasars may be moot since we find
that the colors of quasars are not a very strong function of
luminosity as can be seen in Figure~11 of R01, and because our sample
already includes relatively faint quasars.

If the photometric redshift technique can be calibrated out to
$g'=21$, then we can expect to determine photometric redshifts for as
many as one million quasars \citep{fan99}.  Can we really estimate
redshifts to $21^{\rm st}$ magnitude in the SDSS?  It would seem that
the answer is ``yes'', especially since Figure~\ref{fig:fig4} shows no
systematic deviations from zero at fainter magnitudes.  That is, this
technique is getting incorrect redshifts for as many quasars at
fainter magnitudes as it does at the bright end.

The answer also appears to be ``yes'' from the standpoint of the
photometric limits of the imaging Survey.  The photometric limit of
the Survey in $u'$ is approximately 22.3.  If we consider only UVX
quasars, which have $u'-g' \le 0.8$, then this means that we can use
this technique to $g'=21.5$ (over a magnitude brighter than the Survey
limit in $g'$).  If we cut our sample at $g'=21$, then our sample
should not be affected by incompleteness in $u'$, and the objects will
be sufficiently bright that the errors in their magnitudes will not
adversely affect our results.  Furthermore, we note that the $g'=21$
limit applies only to low-redshift quasars.  For higher redshift
quasars where the Lyman-$\alpha$ forest blocks most of the light from
the quasar in one or more of the SDSS filters, photometric redshifts
can be determined for fainter quasars.

\subsubsection{Efficient Selection of Quasar Candidates}

It is important to mention that finding one million quasars requires
looking at considerably more than one million quasar candidates; not
all SDSS point sources that lie outside of the ``stellar locus'' are
quasars \citep{ric+01}.  All of the objects used in the analysis of
this paper are already known to be quasars.  However, in a real
photometric redshift application, we will not know if all the objects
in the sample are quasars --- in fact we will know that many are {\em
not} quasars.  If the SDSS selection efficiency were as high for faint
quasars as it is for bright quasars (nearly 70\%), then coupled with
our 70\% accuracy of photometric redshifts, we would expect that only
50\% of our proposed faint quasar sample would actually be quasars
with the correct photometric redshift.  Most science applications will
require more than 50\% efficiency/accuracy.  We have already discussed
how we might increase the fraction of correct photo-$z$'s, we now
discuss how to increase the efficiency of quasar selection.

Perhaps the easiest way to increase our selection efficiency is to
consider only small regions of color space.  Without too much
difficulty, it should be possible to pick objects from a restricted
range of color-space such that the vast majority of objects are
quasars.  Even a simple UV-excess cut such as $u'-g'$ $\le 0.6$ will
yield a sample where the majority of point sources will be quasars (if
they are sufficiently faint, see \citealt{fan99}).  Further
color-space restrictions can be used to increase the quasar
probability even more (e.g., color cuts to reduce the contamination
from white dwarfs --- which have SDSS colors that are considerably
different from quasars).

Another approach to increase our efficiency is if non-quasars are
found to contaminate only certain regions of photometric redshift
space.  For example, we find that for a sample of 220 white dwarfs,
135 (75\%) have photometric redshifts between $1.025$ and $1.225$.  If
we were to exclude this redshift region from our analysis, we would
reduce our contamination from white dwarfs considerably.  If other
contaminants were similarly confined in photometric redshift space, we
might be able to increase our efficiency to a point where the
contamination was at an acceptable level.  Of course, this method will
introduce redshift biases in our analysis, but these biases can be
corrected on a statistical basis.

We might also increase the fraction of real quasars in our sample by
looking at the radio properties of the sources.  Point objects that
are radio sources also have a very high probability of being quasars.
If we were to restrict our quasar candidate selection to those object
that are also FIRST sources, we would improve our quasar selection
efficiency considerably, albeit at the cost of a large reduction in
sample size.

Full calibration of the photometric redshift technique will require an
exploration of most of the color parameter space inhabited by quasars,
but individual science applications can take advantage of smaller
regions of color space where the quasar fraction is very high.
Nevertheless, we are confident that quasars can be selected to
sufficient efficiency (or that the efficiency can be modeled
accurately) and their photometric redshifts can be determined to
sufficient accuracy that good science can be done with such samples.
We now turn to a brief discussion of some of these applications.

\section{Science Programs with SDSS Quasar Photometric Redshifts}

The SDSS quasar sample will represent a factor of 100 increase in the
number of quasars from what was recently the largest complete survey
in the published literature, specifically the Large Bright Quasar
Survey (LBQS; \citealt{mwa+91}).  These 100,000 new SDSS quasars,
combined with the 25,000 expected from the 2dF QSO Survey
\citep{boy00,csb+01} will yield over a factor of eight increase in the
number of spectroscopically confirmed quasars known today.  However,
if we can successfully calibrate photometric redshifts to $g'=21$, we
can expect to increase this number by another factor of ten ---
yielding a million or more quasar candidates with photometric
redshifts in the SDSS area.  Although not all of the quasar candidates
will actually be quasars, as we have discussed above, it should be
possible to select subsamples that have a relatively high probability
of being quasars.

There are many scientific uses of a such a large statistical sample of
quasar candidates with photometric redshifts.  Examples of science
that will benefit from photometric redshifts of quasar candidates are:
quasar-galaxy correlations and amplification bias of quasars
\citep[e.g.,][]{twd95,ni99}, quasar pairs and large scale structure as
traced by quasars \citep[e.g.,][]{mwf99,cs96}, the luminosity function
of faint quasars \citep[e.g.,][]{boy00,kk88}, and strong
gravitational lensing \citep{rac+99,ptw+00,mw00}.

Quasar amplification bias research can benefit greatly from
photometric redshifts of quasars.  Current amplification bias studies
are limited by small number statistics, since the number of known
quasars is actually rather small.  A larger sample of quasars would be
a significant aid to this research.  Contamination of the quasar
sample (objects that are not really quasars) merely serves to degrade
the amplification bias signal (whether it be a correlation or an
anti-correlation with galaxies), making a detection using a
photometric redshift sample even more significant.

Strong lensing is another area that is likely to benefit from
photometric redshifts.  If one were merely attempting to find as many
gravitationally lensed quasars as possible, then requiring that pairs
of objects have similar photometric redshifts would increase the
efficiency of such searches.  While the imposition of such a selection
criterion would hinder a statistical analysis to determine
$\Omega_{\Lambda}$, it would be a very efficient way to find lenses
which could be used to determine $H_{\rm o}$.

Quasar-quasar clustering studies also stand to gain from such a
photo-$z$ analysis.  Even though the SDSS will confirm upwards of
$10^5$ quasars, quasar-quasar clustering studies require that we take
into account the SDSS's minimum fiber separation limit and the extent
to which spectroscopic plates overlap each other.  Photometric
redshifts of quasars can be used to study clustering to fainter limits
and higher density than the spectroscopic survey --- without the need
for corrections due to spatial selection effects.  More importantly,
photometric redshifts of quasars will help greatly by increasing the
overall density of object that can be used in quasars clustering
studies; this is important because of the relative sparseness of
bright quasars (as compared to normal galaxies).

Finally we discuss the luminosity function of faint quasars.  The
photometric redshift technique is not very useful unless it can be
used to fainter limits than the spectroscopic part of the Sloan
Survey.  The calibration of the method fainter than $i'=19$ will
require an extensive follow-up campaign.  Such a calibration sample
will be interesting in and of itself.  These faint calibration quasars
and the eventual faint photo-$z$ quasars can be used to determine the
quasar luminosity function at faint limits, which is of great interest
in terms of studying the evolution of quasars.

For those programs that are not adversely affected by moderate
contamination (i.e., many of the objects really are not quasars) and
errors in the redshifts, large regions of color space can be used to
select quasar candidates.  In this case, the SDSS imaging data can be
thought of as an objective prism survey.  For more sensitive programs,
smaller regions of parameter space (e.g., those that are not
contaminated by white dwarfs) can be defined that will give less
contamination and more accurate redshifts at the cost of biasing the
redshift distribution.

In addition to contamination by non-quasars, redshift errors for real
quasars will also need to be accounted for.  Although an error in
redshift as large as 0.2 may seem large for those who have experience
with photometric redshifts of galaxies, for quasars an error of 0.2 is
adequate for most science applications.  For example, an error of 0.2
is sufficiently small to allow one to split a photo-$z$ sample into
``high-redshift'' and ``low-redshift''.  We also emphasize that, to
the extent that the photo-$z$ method can be considered low-resolution
spectroscopy, a redshift error of $\Delta z = 0.2$ is really quite
good.  For example, at $z=3$ the wavelength error of Lyman-$\alpha$
due to an error in redshift of $\Delta z = 0.2$ would only be
$243\,{\rm \AA}$, which is smaller than the separation between the
Lyman-$\alpha$ and \ion{Si}{4} emission lines.

\section{Conclusions}

The color-redshift relation of quasars has considerable structure as
seen through the SDSS filters.  This structure can be used in four
dimensions to break the one dimensional degeneracies between quasar
colors and their redshifts, and allows for the determination of
photometric redshifts, including low-redshift quasars ($z\le2.2$).
While it is clear that the structure induced by the Lyman-$\alpha$
forest allows for accurate determination of quasar redshifts at large
redshift ($z\ge3.0$), to our knowledge this work, together with
\citet{bud01}, is the first successful attempt to determine
photometric redshifts for a large sample of low-redshift quasars using
as few as four broad-band colors.

We can expect at least 70\% of the photometric redshifts to be
accurate to $|\Delta z| = 0.2$ for quasars with $0 < z < 5$, and this
fraction is not apparently a function of magnitude (to at least
$i'=20$, if not fainter).  With a limiting magnitude of $u'=22.3$, the
SDSS should be able to determine photometric redshifts for quasars
brighter than about $g'=21.0$, which yields on the order of one
million quasar candidates over the entire Survey area.  Photometric
redshifts for high-redshift quasars can be determined to even fainter
limits.

We advocate the construction of a large sample of quasar spectra
($\sim1000$) fainter than $i'=19$ in order to calibrate the
photometric relation for magnitudes fainter than the limiting
magnitude of the main SDSS quasar survey.  The construction of such a
sample may allow for the eventual determination of redshifts for a
million or more quasar candidates through the photometric redshift
techniques described herein.  This sample will also be useful for
studying the faint end of the quasar luminosity function.

Although one million quasar candidates with photometric redshifts are
by no means equivalent to one million quasars with spectroscopic
redshifts, there are many scientific topics that can be addressed with
such a sample.  Some science is relatively insensitive to the quasar
selection efficiency and to errors in the redshifts of those objects
that are quasars; other programs will require detailed studies of the
selection effects and redshift errors in order to take full advantage
of the sample.  With a bit of effort, it should be possible to create
smaller subsamples of the nearly one million quasar candidates that
have both a high probablitiy of being a quasar and an accurate
photometric redshift.  In any case, we expect that there will be no
lack of creative uses for the sample of quasar candidates with
photometric redshifts that the SDSS will produce.

\acknowledgements

The Sloan Digital Sky Survey\footnote{The SDSS Web site is
http://www.sdss.org/.} (SDSS) is a joint project of The University of
Chicago, Fermilab, the Institute for Advanced Study, the Japan
Participation Group, The Johns Hopkins University, the
Max-Planck-Institute for Astronomy (MPIA), the Max-Planck-Institute
for Astrophysics (MPA), New Mexico State University, Princeton
University, the United States Naval Observatory, and the University of
Washington. Apache Point Observatory, site of the SDSS telescopes, is
operated by the Astrophysical Research Consortium (ARC).  Funding for
the project has been provided by the Alfred P. Sloan Foundation, the
SDSS member institutions, the National Aeronautics and Space
Administration, the National Science Foundation, the U.S. Department
of Energy, the Japanese Monbukagakusho, and the Max Planck Society.
This research has made use of the NASA/IPAC Extragalactic Database
(NED) which is operated by the Jet Propulsion Laboratory, California
Institute of Technology, under contract with the National Aeronautics
and Space Administration.  GTR, MAW, and DPS acknowledge support from
NSF grant AST99-00703.  XF was supported in part by NSF grant
PHY00-70928 and a Frank and Peggy Taplin Fellowship.  MAS was
supported in part by NSF grant AST00-71091.  XF and MAS further
acknowledge support from the Princeton University Research Board and a
Porter O. Jacobus Fellowship.  We thank David Weinberg for his
comments on the manuscript.  We also thank an anonymous referee for a
number of constructive suggestions.

\clearpage

\end{document}